\documentclass[sigconf,natbib=false]{acmart}

\copyrightyear{2026}
\acmYear{2026}
\setcopyright{cc}
\setcctype{by}
\acmConference[Q-SE '26]{7th International Workshop on Quantum Software Engineering }{April 12--18, 2026}{Rio de Janeiro, Brazil}
\acmBooktitle{7th International Workshop on Quantum Software Engineering (Q-SE '26), April 12--18, 2026, Rio de Janeiro, Brazil}
\acmPrice{}
\acmDOI{10.1145/3786150.3788611}
\acmISBN{979-8-4007-2383-4/2026/04}
\usepackage[inline]{enumitem}
\usepackage{glossaries-extra}
\usepackage{xspace}
\usepackage{amsmath}
\usepackage{braket}
\usepackage{csquotes}
\usepackage{orcidlink}
\usepackage[shortcuts]{extdash}

\RequirePackage[
  datamodel=acmdatamodel,
  style=acmnumeric,
  backend=biber,
  maxcitenames=3,
  minbibnames=1,
  mincitenames=1,
  maxbibnames=3,
  uniquelist=false,
  giveninits=true,
  url=false,
  doi=false,
  eprint=false,
  isbn=false,
  ]{biblatex}

\newcommand{\eg}{\emph{e.g.}\xspace}
\newcommand{\ie}{\emph{i.e.}\xspace}
\newcommand{\cf}{\emph{cf.}\xspace}
\newcommand{\etal}{\emph{et al.}\xspace}

\newcommand{\giturl}{https://github.com/cirKITers/effect-of-noise-in-qfms}
\newcommand{\zenodourl}{https://doi.org/10.5281/zenodo.15211318}

\usepackage{pifont}
\makeatletter
\newcommand{\linebreakand}{%
  \end{@IEEEauthorhalign}
  \hfill\mbox{}\par
  \mbox{}\hfill\begin{@IEEEauthorhalign}
}
\makeatother

\definecolor{lfd1}{HTML}{FFFFFF} 
\definecolor{lfd2}{HTML}{E69F00}
\definecolor{lfd3}{HTML}{999999}
\definecolor{lfd4}{HTML}{009371}

\definecolor{lfd5}{HTML}{BEAED4}
\definecolor{lfd6}{HTML}{ED665A}
\definecolor{lfd7}{HTML}{1F78B4}

\setabbreviationstyle[acronym]{long-short}
\glssetcategoryattribute{acronym}{nohyperfirst}{true}

\newacronym{nisq}{NISQ}{noisy intermediate-scale quantum}
\newacronym{ftqc}{FTQC}{fault-tolerant quantum computing}
\newacronym{qml}{QML}{quantum machine learning}
\newacronym{qpu}{QPU}{quantum processing unit}
\newacronym{qcd}{QCD}{quantum chromodymamics}
\newacronym{qed}{QED}{quantum electrodynamics}
\newacronym{mde}{MDE}{model-driven engineering}
\newacronym{dsml}{DSML}{domain-specific modelling language}
\newacronym{qsvt}{QSVT}{quantum singular value transformation}
\newacronym{se}{SE}{software engineering}
\newacronym[\glslongpluralkey={Lattice Gauge Theories}]{lgt}{LGT}{Lattice Gauge Theory}

\usepackage{tikz}
\usepackage{yquant}
\usepackage{tcolorbox}
\tcbuselibrary{skins,breakable}
\usetikzlibrary{positioning, calc, arrows.meta, shadows.blur, intersections, fit, math}

\newlength{\WIDTH}\newlength{\HEIGHT}

\definecolor{lfd1}{HTML}{000000} 
\definecolor{lfd2}{HTML}{E69F00}
\definecolor{lfd3}{HTML}{999999}
\definecolor{lfd4}{HTML}{009371}
\definecolor{lfd5}{HTML}{BEAED4}
\definecolor{lfd6}{HTML}{ED665A}
\definecolor{lfd7}{HTML}{1F78B4}
\definecolor{lfd8}{HTML}{009371}

\def\particlesize{4mm}
\def\atomsize{1.5mm}
\def\offset{4mm}
\def\textoffset{0.6pt}
\def\atomoffset{1pt}
\def\stepdist{3mm}
\def\innerdist{5mm}
\def\particledist{1.15cm}
\def\uodd{11mm}
\def\ueven{9mm}
\def\umid{3mm}
\def\atomheight{6.5mm}
\def\titleboxwidth{7mm}
\def\abstractionwidth{2.2cm}
\def\abstractionheight{0.1cm}
\def\abstoffset{1.8cm}
\def\boxdist{2mm}
\def\boxwidth{0.3\columnwidth}
\def\boxheight{0.5cm}
\def\bbwidth{4mm}
\def\bbheight{6mm}
\def\hamiltonianheight{0.8cm}
\def\hamiltonianoffset{0cm}
\def\tedist{1cm}
\def\timeevolutionwidth{0.5*\columnwidth-0.8\tedist}
\def\teinnerdist{2mm}
\def\trotterwidth{0.07*\columnwidth}
\def\trotterspace{0.01*\columnwidth}
\def\hamiltonianwidth{1.5cm}
\def\unitaryheight{6.5mm}

\tikzset{linkarr/.style={-{Triangle[length=2mm, scale width=0.5]}, draw=lfd3, line width=1.5mm}}
\tikzset{blockarr/.style={-{Triangle[length=3.5mm, scale width=0.5]}, draw=lfd2, line width=3.5mm}}
\tikzset{blockline/.style={-, draw=lfd2, line width=3.5mm}}
\tikzset{doubleblockarr/.style={{Triangle[length=3.5mm, scale width=0.5]}-{Triangle[length=3.5mm, scale width=0.5]}, draw=lfd3, line width=3.5mm}}
\tikzset{blockarrgray/.style={-{Triangle[length=3mm, scale width=0.45]}, draw=lfd3, line width=2.5mm}}
\tikzset{doubleblockarrgray/.style={{Triangle[length=3mm, scale width=0.45]}-{Triangle[length=3mm, scale width=0.45]}, draw=lfd3, line width=2.5mm}}
\tikzset{bbarr/.style={-{Triangle[length=3mm, scale width=0.6]}, draw=lfd3, line width=0.8mm, dotted}}
\tikzset{bbline/.style={draw=lfd1, line width=0.8mm, dotted}}
\tikzset{desc/.style={font=\scriptsize}}
\tikzset{blockdesc/.style={font=\small}}
\tikzset{title/.style={rectangle, font=\large, fill=lfd4!20, draw=lfd4, minimum width=1.5cm, minimum height=1cm}}
\tikzset{bb/.style={rectangle, font=\tiny, fill=lfd4!20, draw=lfd4, minimum width=\bbwidth, minimum height=\bbheight, align=center}}
\tikzset{titletext/.style={font=\small, text=lfd4,align=center,rotate=90}}
\tikzset{abstext/.style={font=\scriptsize, text=lfd2,align=center}}
\tikzset{titlebox/.style={rectangle, font=\large, fill=lfd4!20, draw=lfd4, minimum width=\titleboxwidth}}
\tikzset{abstractbox/.style={rectangle, font=\small, fill=lfd2, draw=lfd2, minimum width=\abstractionwidth, blur shadow, rounded corners, text=white,align=center}}
\tikzset{smallabstractbox/.style={inner sep = 2pt,font=\scriptsize, minimum width=0.5*\abstractionwidth, minimum height=\abstractionheight, rounded corners, text=white,align=center,fill=lfd3}}
\tikzset{atom/.style={circle, fill=lfd4, draw=lfd4, minimum size=\atomsize, inner sep=0pt}}
\tikzset{transparent/.style={opacity=0.7, fill opacity=0.7}}
\tikzset{unitary/.style={draw=lfd1, fill=lfd1!20}}
\tikzset{unitaryblock/.style={draw=lfd2, fill=lfd2, rounded corners=1pt, text=white, minimum height=\unitaryheight}}
\tikzset{box/.style={rectangle, fill=lfd3!20, draw=lfd1, minimum height=\boxheight, minimum width=\boxwidth}}
\tikzset{hamiltonian/.style={rectangle, font=\normalsize, fill=lfd4, draw=lfd4, minimum width=\hamiltonianwidth, minimum height=\hamiltonianheight, blur shadow, rounded corners=1pt, text=white,align=center}}
\tikzset{smallhamiltonian/.style={font=\tiny, lfd4, minimum width=0.1*\hamiltonianwidth, minimum height=0.1*\hamiltonianheight, text=lfd4,align=center, inner sep=0}}
\tikzset{hamiltonianarr/.style={{Triangle[length=2mm, scale width=0.6]}-{Triangle[length=2mm, scale width=0.6]}, draw=lfd3, line width=0.8mm}}
\tikzset{timeevolution/.style={|-{Triangle[length=1.5mm, scale width=0.6]}, draw=lfd1, line width=0.3mm}}

\pgfdeclarelayer{bg}
\pgfdeclarelayer{bg1}
\pgfdeclarelayer{bg2}
\pgfdeclarelayer{bg3}
\pgfsetlayers{bg3,bg2,bg1,bg,main}

\hypersetup{colorlinks=true,urlcolor=teal,
            linkcolor=teal,citecolor=teal}

\begin{document}

\title{Towards Quantum Software for Quantum Simulation}

\author{Maja Franz\orcidlink{https://orcid.org/0000-0002-2801-7192}, Lukas~Schmidbauer\orcidlink{0009-0001-7171-0865}}
\email{{maja.franz, lukas.schmidbauer}@othr.de}
\affiliation{%
  \institution{Technical University of Applied Sciences Regensburg}
  \city{Regensburg}
  \country{Germany}
}
\author{Joshua Ammermann\orcidlink{https://orcid.org/0000-0001-5533-7274}, Ina~Schaefer\orcidlink{0000-0002-7153-761X}}
\email{{joshua.ammermann, ina.schaefer}@kit.edu}
\affiliation{%
  \institution{Karlsruhe Institute of Technology}
  \city{Karlsruhe}
  \country{Germany}
}
\author{Wolfgang Mauerer\orcidlink{0000-0002-9765-8313}}
\email{wolfgang.mauerer@othr.de}
\affiliation{%
  \institution{Technical University of Applied Sciences Regensburg/Siemens AG, Foundational Technologies}
  \city{Regensburg/Munich}
  \country{Germany}
}

\begin{abstract}
  Quantum simulation is a leading candidate for demonstrating practical quantum advantage over classical computation, as it is believed to provide exponentially more compute power than any classical system.
  It offers new means of studying the behaviour of complex
  physical systems, for which conventionally software-intensive simulation codes based on numerical high-performance computing are used.
  Instead, quantum simulations map properties and characteristics of subject systems, for instance chemical molecules, onto quantum devices that then mimic the system under study.
  
  Currently, the use of these techniques is largely limited to fundamental science, as the overall approach remains tailored for specific problems: We lack infrastructure and modelling abstractions that are provided by the \gls{se} community for other computational domains.
  
  In this paper, we identify critical gaps in the quantum simulation software stack~--~particularly the absence of general-purpose frameworks for model specification, Hamiltonian construction, and hardware-aware mappings.
  We advocate for a modular \gls{mde} approach that supports different types of quantum simulation (digital and analogue), and facilitates automation, performance evaluation, and reusability.
  Through an example from high-energy physics, we outline a vision for a quantum simulation framework capable of supporting scalable, cross-platform simulation workflows.
\end{abstract}


\maketitle

\section{Introduction}

Quantum simulation is widely regarded as one of the most promising near-term applications of quantum computing~\cite{snowmass23_simulation, Miessen_2022, altman21, franz24_chep}.
Conceptionally, quantum simulation is similar to a small-scale model of an aircraft in a wind tunnel:
instead of utilising intricate algorithms and numerically solving the fluid-dynamical equations of motion, the model itself experiences the phenomena under study~\cite{hangleiter22}.
Similarly, utilising the dynamical capabilities of a programmable quantum system (in other words, a quantum computer) to learn about the dynamics of an appropriately mapped subject system, offers a path to tackle problems that are classically intractable~\cite{meurice22,troyer05}.

However, the \emph{software infrastructure} and modelling frameworks required to systematically develop, deploy, and analyse these simulations remain significantly underdeveloped, as we discuss in this work.
The state-of-the-art of quantum simulation is largely model-based:
A subject Hamiltonian, which describes the dynamics of a physical target system gets mapped onto quantum computers by manually constructing models to approximate the target dynamics (see, \eg,~Refs.~\cite{yang20, zhou22}).
This approach is not only central to physics-oriented applications but also underpins the potential for exponential quantum advantage in a broad class of algorithmic primitives based on \gls{qsvt}~\cite{Dong_2022}.
However, in contrast to quantum optimisation or quantum machine learning, where well-established algebraic modelling techniques and \glspl{dsml} enable \textbf{automated transformation} into quantum\-/executable formats~\cite{lucas14, strobl:25:qml_essentials,schmidbauer24, schmidbauer25}, the process of preparing a quantum simulation remains largely manual, problem-specific, and hardware-dependent.
The absence of a generic software stack severely limits the ability to explore the space of quantum simulation strategies,
which makes it difficult to \textbf{generalise implementations}, quantify trade-offs between alternative mappings, or benchmark against classical methods.
Additionally, the challenges are exacerbated by intricacies of quantum hardware that require substantial customisation~\cite{snowmass23_simulation}.

Furthermore, when considering classical simulations of physical systems, empirical benchmarking and performance analysis are routine practices~\cite{daley22}.
However, for quantum systems such assessments are hindered by the lack of robust \textbf{simulation and benchmarking tooling}.
Existing classical methods such as Monte Carlo simulations face challenges like the sign problem~\cite{troyer05}, yet they benefit from decades of software development.
In contrast, quantum simulations~--~though theoretically immune to certain classical limitations~--~often lack the practical scaffolding to exploit this advantage meaningfully.

To move beyond isolated demonstrations, the field must adopt a more \textbf{systematic and modular} approach to simulation design~\cite{di_matteo25}.
Specifically, it is necessary to:
\begin{enumerate*}[label=(\arabic*)]
    \item formalise the construction of effective model Hamiltonians from physical theories~\cite{meurice22},
    \item automate the mapping of these models to quantum hardware representations, and
    \item develop standardised simulation workflows that allow for scalable execution, validation, and performance estimation.
\end{enumerate*}
Building on the general quantum \gls{se} roadmap presented in Ref.~\cite{murillo25}, this work outlines the initial stages of developing a quantum simulation software framework, an area that has received limited attention in previous studies.
We identify key stages in a simulation pipeline, from model specification and Hamiltonian design to hardware compilation, and explore open challenges in the pathway to systematic software support.
Ultimately, enabling a software stack for quantum simulation~--~including domain-specific abstractions, modelling languages, and tooling for mapping and optimisation~--~is critical for unlocking the full potential of this physics-driven quantum approach.
In the following, we outline the fundamentals of quantum simulation in \autoref{sec:sim-methods}.
\autoref{sec:software_framework} then describes our vision of a general quantum simulation framework at an example from the high-energy physics domain, which aims at a \gls{mde}~\cite{france07} approach to quantum simulations.
\autoref{sec:discussion} lists open challenges.

\section{Quantum Simulation}
\label{sec:sim-methods}
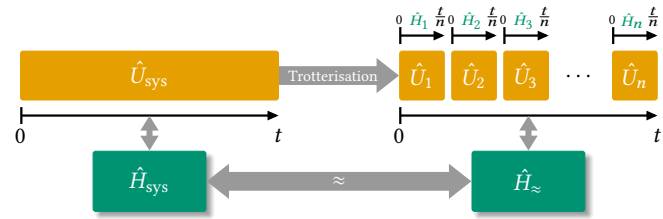
\begin{figure}[h]
  \centering
  \setlength{\WIDTH}{\columnwidth}
\setlength{\HEIGHT}{7cm}

\begin{tikzpicture}
  \coordinate (tl) at (0, \HEIGHT);
  \coordinate (tm) at (0.5*\WIDTH, \HEIGHT);
  \coordinate (tr) at (\WIDTH, \HEIGHT);
  \coordinate (bl) at (0, 0);
  \coordinate (br) at (\WIDTH, 0);

  \coordinate (usysstart) at ($(tl) + (0, -\hamiltonianoffset)$);
  \coordinate (usysend) at ($(tm) + (-0.8*\tedist, -\hamiltonianoffset)$);
  \coordinate (usysmid) at ($(usysstart)!0.5!(usysend)$);
  \draw[timeevolution] (usysstart) -- (usysend);
  \node[anchor=north, yshift=-0.3*\teinnerdist] at (usysstart) {$0$};
  \node[anchor=north, yshift=-0.3*\teinnerdist] at (usysend) {$t$};
  \node[unitaryblock, anchor=south, minimum width=\timeevolutionwidth, yshift=\teinnerdist] (usys) at ($(usysstart)!0.5!(usysend)$) {$\hat{U}_\text{sys}$};

  \coordinate (uapproxstart) at ($(tm) + (0.8*\tedist, -\hamiltonianoffset)$);
  \coordinate (uapproxend) at ($(tr) + (0, -\hamiltonianoffset)$);
  \draw[timeevolution] (uapproxstart) -- (uapproxend);
  \node[anchor=north, yshift=-0.3*\teinnerdist] at (uapproxstart) {$0$};
  \node[anchor=north, yshift=-0.3*\teinnerdist] at (uapproxend) {$t$};

  \node[hamiltonian, anchor=north,yshift=-\tedist] (hsys) at (usys) {$\hat{H}_\text{sys}$};
  \coordinate (uapproxmid) at ($(uapproxstart)!0.5!(uapproxend)$);
  \node[hamiltonian] (happrox) at (hsys-|uapproxmid) {$\hat{H}_\approx$};
  \node[unitaryblock, anchor=south west, minimum width=\trotterwidth, yshift=\teinnerdist] (u1) at (uapproxstart) {$\hat{U}_1$};
  \node[unitaryblock, anchor=west, minimum width=\trotterwidth] (u2) at ($(u1.east) + (\trotterspace, 0)$) {$\hat{U}_2$};
  \node[unitaryblock, anchor=west, minimum width=\trotterwidth] (u3) at ($(u2.east) + (\trotterspace, 0)$) {$\hat{U}_3$};
  \node[unitaryblock, anchor=south east, minimum width=\trotterwidth, yshift=\teinnerdist] (un) at (uapproxend) {$\hat{U}_n$};
  \node[] at ($(u3)!0.5!(un)$) {$\dots$};

  \foreach \i in {1,...,3}
  {
    \coordinate (uapproxstartup\i) at ($(u\i.north west) + (0, \teinnerdist)$);
    \coordinate (uapproxendup\i) at ($(u\i.north east) + (0, \teinnerdist)$);
    \draw[timeevolution] (uapproxstartup\i) -- node[pos=0.45,smallhamiltonian,yshift=\teinnerdist] {$\hat{H}_\i$}(uapproxendup\i);
    \node[font=\tiny,anchor=south, yshift=0.2*\teinnerdist] at (uapproxstartup\i) {$0$};
    \node[font=\tiny,anchor=south, xshift=-0.3*\teinnerdist, yshift=0.1*\teinnerdist] at (uapproxendup\i) {$\frac{t}{n}$};
  }

  \coordinate (uapproxstartupn) at ($(un.north west) + (0, \teinnerdist)$);
  \coordinate (uapproxendupn) at ($(un.north east) + (0, \teinnerdist)$);
  \draw[timeevolution] (uapproxstartupn) -- node[pos=0.45,smallhamiltonian,yshift=\teinnerdist] {$\hat{H}_n$}(uapproxendupn);
  \node[font=\tiny,anchor=south, yshift=0.2*\teinnerdist] at (uapproxstartupn) {$0$};
  \node[font=\tiny,anchor=south, xshift=-0.3*\teinnerdist, yshift=0.1*\teinnerdist] at (uapproxendupn) {$\frac{t}{n}$};

  \draw[hamiltonianarr] (hsys) -- (usysmid);
  \draw[hamiltonianarr] (happrox) -- (uapproxmid);
  \draw[doubleblockarrgray,font=\small] ($(hsys.east) + (0*\teinnerdist,0)$) -- node[midway,text=white] {$\approx$} ($(happrox.west) - (0*\teinnerdist,0)$);
  \draw[blockarrgray] ($(usys.east) + (0*\teinnerdist,0)$) -- node[pos=0.45,text=white,font=\scriptsize] {Trotterisation} ($(u1.west) - (0*\teinnerdist,0)$);

\end{tikzpicture}
  \caption{Continuous vs. discrete (Trotterised) time evolution. Left: The continuous dynamic of a physical systems is described by the Hamiltonian $\hat{H}_\text{sys}$ starting at an initial state at time $0$, which then evolves under the corresponding unitary $\hat{U}_\text{sys}$ to a target state at time $t$. Right: Continuous time evolution is discretised (\emph{Trotterised}) into local one- and two-qubit gates that can be executed on universal, gate-based quantum hardware, yet in the general case results in a time evolution under an approximated Hamiltonian $\hat{H}_\approx$.}
  \label{fig:time_evolution}
\end{figure}

The (classically hard to solve) time evolution of (quantum) systems (\eg, wind tunnel, electron interactions) is governed by a Hamiltonian $\hat{H}_\text{sys}$ that contains all information about the physics of that system.
Starting at the systems initial state, configurable quantum simulators (\ie, quantum \enquote{computers}) can mimic this time evolution~\cite{lamata18, yeung24}, expressed by the unitary operator
\(
  \hat{U}_\text{sys}(t) = \exp\left(-i \int_{0}^{t}\hat{H}_\text{sys}(t)\text{d}t\right).
\)
Importantly, the equation connects unitary gates \(\hat{U}_\text{sys}\), the usual objects of study in quantum computing, with Hamiltonians \(\hat{H}_\text{sys}\) (see left part of \autoref{fig:time_evolution}).
We discuss two principal approaches (\ie, digital and analogue) that realise this time evolution in the following (see \citeauthor{daley22}~\cite{daley22} for a detailed discussion).

\subsection{Digital Quantum Simulation}
For many physical systems, $\hat{H}_\text{sys}$ can be expressed as a sum of (non-commuting) $k_j$-local interaction terms: $\hat{H}_\text{sys} = \sum_j \hat{H}_j$.
To address the non-commutativity in $\hat{H}_j$, a \emph{Trotterisation} applies the $k_j$-local unitaries $\hat{U}_j = e^{-i\hat{H}_j(t/n)}$ $n$ times over time slices $t/n$:
\vspace*{-0.5em}\begin{equation}
  \hat{U}_\text{sys}(t) = e^{-i \sum_j \hat{H}_j(t)} = \underbrace{\left( \prod_j e^{-i \hat{H}_j(t/n)} \right)^n}_{\hat{U}_\approx(t) = \exp(-i \int_{0}^{t}\hat{H}_\approx(t)\text{d}t)} +\, \underbrace{\mathcal{O}\left(\frac{t^2}{n}\right),}_{\text{\clap{\vbox{\hbox{Approximation}\hbox{ error}}}}}\vspace*{-0.25em}
\end{equation}
where $\hat{U}_\approx$ approximates time evolution under an implicit Hamiltonian $\hat{H}_\approx$ (right part of \autoref{fig:time_evolution}).
This decouples executing hardware from simulated dynamics for error-corrected systems, and can implement a large class of Hamiltonians (see Ref.~\cite{halimeh2025} for examples), although discretisation leads to approximation errors and poorer scalability.
Digital quantum simulation is universal as it allows for  approximating arbitrary single- and two-qubit gates to arbitrary accuracy.
This enables simulating the dynamics of quantum systems that may be fundamentally different from the simulating hardware.
The resulting unitary gate instructions can be formulated using existing \glspl{dsml}~\cite{openqasm, qiskit, cirq} that can be executed on gate-based quantum computers.
As is common in the circuit model, a large number of high-fidelity qubits and deep circuits are typically required.
Error mitigation techniques allow for limited execution on \gls{nisq} machines, but not yet at scale and precision needed for practical relevance~\cite{lamata18}.

\subsection{Analogue Quantum Simulation}
In contrast to digital simulation, analogue quantum simulation~\cite{hangleiter22} aims to closely mimic the characteristics of the simulated quantum model by evolving under a simulator Hamiltonian $\hat{H}_\text{sim}$ that is similar to the simulated system Hamiltonian $\hat{H}_\text{sys}$.
By directly mapping $\hat{H}_\text{sys} \mapsto \hat{H}_\text{sim}$ the time evolution proceeds under the natural Hamiltonian evolution of the simulator Hamiltonian and corresponding hardware.
In particular in the context of physics-related questions, such as applications from high-energy physics, analogue quantum simulation seen as proof of (often exponential) quantum advantages over the best known classical techniques~\cite{Bermejo_Vega_2018,daley22, Miessen_2022}.
The direct implementation of $\hat{H}_\text{sim}$ on a quantum system is promising in terms of scalability, but also entails limitations due to physical restrictions and non-universality of operations~\cite{yeung24} that restrict flexibility.
When $\hat{H}_\text{sys}$ is structurally similar to $\hat{H}_\text{sim}$, efficient mappings can be determined comparatively easily, while it remains a major challenge for more general problems.
Although efforts in digital simulation have started to explore general algorithmic approaches~\cite{shaw20}, analogue simulations still rely heavily on problem-specific mappings that are hard to scale or reuse.
Analogue simulation lacks modelling tools and formal abstractions that have accelerated progress in other areas, typically based on ideas from software engineering.

Lamata~\etal~\cite{lamata18} suggest a mixed approach where analogue blocks provide scalability by reducing gate-counts and hence sources of noise, while digital steps amplify the variety of possible operations.
This may be useful on the way towards error-correcting codes to achieve usable results before perfect hardware is available.

\section{Vision: Quantum Simulation Framework}
\label{sec:software_framework}

\begin{figure}[htbp]
  \centering
  \input{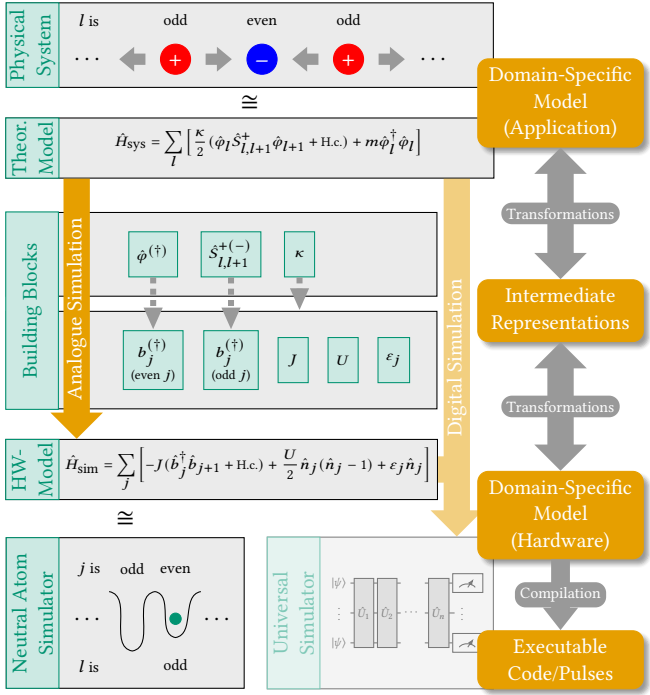}
  \caption{Envisioned quantum simulation framework illustrated for one-dimensional \acrlong{qed} (left),
  and suggested software abstraction layers (right).
  }\vspace*{-2em}
  \label{fig:sim_expl}
\end{figure}

Current quantum simulations often require manual processes for translating models into executable programs on quantum hardware.
Instead, we propose a \gls{mde}~\cite{france07} approach to quantum simulation, where high-level, abstract models guide automated transformations from theory to implementation.
Using a concrete example, in which a theory from high-energy physics is simulated in an analogue mode, we highlight both the demands on quantum software and the deficiencies in today's tooling.
As an instance of the software stack presented in Ref.~\cite{ammermann:24:towards}, \autoref{fig:sim_expl} illustrates our envisioned quantum simulation stack, tracing the transformation from the physical model to its realisation on configurable quantum hardware.

\subsection{Example: Simulating a Lattice Gauge Theory}
In our example we demonstrate the software requirements for simulating a physical system~\cite{shaw20}, using an array of neutral atoms in an optical lattice as a simulation platform.
This follows the simulation methodologies developed in~\cite{yang20, zhou22}.

A mathematical description of the target physical system, the \textbf{theoretical model}, captures the physical processes  governing the system.
In our case, we consider a toy model for one-dimensional \acrlong{qed} in a spatial lattice.
While we cannot provide comprehensive physical details (see, \eg, Ref.~\cite{shaw20}), suffice it to say the Hamiltonian $\hat{H}_\text{sys}$ (\cf \autoref{fig:sim_expl}) comprises (a) matter field operators $\hat{\varphi}_l^{(\dagger)}$, where $\hat{\varphi}_l^{\dagger}$ creates a Fermion and  \(\hat{\varphi}_l\) annihilates one at lattice site $l$ with mass $m$, and (b) gauge fields described by spin-$\frac{1}{2}$ operators $\hat{S}_{l, l+1}^{+(-)}$ on the links between neighbouring sites. Local Gauss law constraints, enforced via coupling $\kappa$, ensure physical constraints.
All ingredients are tied to the underlying physics, and share little commonality with computer science thinking.

For execution, Hamiltonian $\hat{H}_\text{sys}$ is mapped to Hamiltonian $\hat{H}_\text{sim}$ (\cf \autoref{fig:sim_expl}) governed by a \textbf{hardware model} (\ie, in this case the \emph{Bose-Hubbard} model), which can experimentally be realised with an array of neutral atoms.
Term \(J\) in $\hat{H}_\text{sim}$ describes the \enquote{hopping} or tunnelling of atoms between adjacent lattice sites: $\hat{b}_{j+1}$ removes an atom at site \(j+1\), and $\hat{b}_j^{\dagger}$ creates one at site \(j\).
The term leading with $U$ represents the strength of the
on-site interaction with $\hat{n}_j = \hat{b}_j^\dagger \hat{b}_j$, which is relevant when more than one atom is located on a single site. The $\varepsilon_j$ term represents a energy offset to represent lattice potentials and suppress long\-/range tunnelling along the 1D chain.

In a \textbf{neutral atom simulator} model parameters can be tuned through laser intensity or magnetic/optical fields~\cite{deutsch00}, such that the simulator Hamiltonian $\hat{H}_\text{sim}$ closely approximates $\hat{H}_\text{sys}$, preserving physical local constraints
~\cite{yang20, zhou22}.
A parity encoding can map the target model onto the simulator (\cf lower left part of \autoref{fig:sim_expl}):
\begin{enumerate*}[label=(\arabic*)]
  \item even-indexed sites represent Fermionic matter fields $\hat{\varphi}_l$ and
  \item odd ones gauge field links via spin operators.
\end{enumerate*}
Hamiltonian \(\hat{H}_{\text{sim}}\) is the lowest level of abstraction from a user's point of view, which however must be further transpiled into control parameters and pulses for execution.
While digital gates are firmly established in frameworks~\cite{qiskit, cirq, openqasm}, pulse-level abstractions~\cite{pasqal, openqasm, lobser23} remain evanescent, depend on hardware-specifics, and are not widely supported by hardware vendors.
Work on analogue quantum simulation remains largely experimental, and lacks intermediate representations of Hamiltonians that can be transformed into pulse-level representations~\cite{ramsauer2025}.
This suggests direct access to executing systems is required in the sense of \emph{individual experiments}, rather than quantum \emph{computers}.

\paragraph{Alternative Simulation Realisations}
Analogue simulation is tightly coupled to the experimental platform, and requires bespoke configuration of physical parameters without enjoying support from software abstractions.
Multiple mappings of the Bose-Hubbard model to other platform have been proposed, for instance trapped ions~\cite{hauke13} or digital simulation via Trotterisation~\cite{shaw20} that approximates continuous dynamics using gate-based circuits.
This highlights diversity of platforms and methods, and  the absence of a unified software framework capable of targeting multiple backends from a single high-level model.
A corresponding interface and the integration with programmable quantum simulators are open challenges.

\subsection{Need for Automation and Abstraction}
Translating theoretical building blocks into a quantum simulation pipeline presents significant complexity
and calls for automated, reusable, and programmable solutions. This motivates the use of an \gls{mde} approach and software abstractions that go
 beyond extensions of traditional approaches like 
 UML~\cite{Yue:2023}. While the roadmap by \citeauthor{murillo25}~\cite{murillo25} recognises the need for \gls{mde} in quantum \gls{se}, 
it does not discuss quantum simulation.
Following \citeauthor{Carbonelli:2024}~\cite{Carbonelli:2024}, we argue these challenges are particularly relevant in the simulation domain, and see need to address the following:

\paragraph{Domain-specific abstractions of physical theories}
To effectively model quantum systems, it is essential to capture the underlying physical theories—such as Hamiltonian dynamics or many-body interactions—through high-level, domain-specific abstractions. These abstractions serve as a bridge between theoretical formulations and their executable representations, enabling automation and reuse.

\paragraph{Scalable, hardware-independent intermediate representations}
A key requirement for portability and scalability in quantum simulation is the use of intermediate representations that decouple the simulation logic from specific hardware architectures. These representations must be expressive enough to encode complex quantum operations while remaining amenable to optimization and analysis.

\paragraph{Code generation and hardware-specific mappings}
Automated code generation tools must translate abstract models into efficient, hardware-compatible code, accounting for unique constraints and capabilities of quantum backends. This involves optimising for gate sets, qubit connectivity, and target hardware characteristics.

\paragraph{Methods for transforming between abstractions}
Consistency across abstraction levels requires formal, well-defined transformations to ensure changes at one level
propagate correctly to others,
and preserve semantic integrity.
We advocate for \glspl{dsml} as
model representations.
Benefits extend beyond physics-centric domains, as quantum simulation plays a role in other domains (\eg, wind tunnel testing or computational fluid dynamics~\cite{fresca21}). 

\section{Open Challenges}
\label{sec:discussion}
Quantum simulation workflows are often ad hoc, and tailored to specific platforms.
A significant gap remains in the systematic translation of high-level theoretical models into executable intermediate representations compatible with constraints and features of quantum hardware.
This hinders general-purpose solutions, and raises research questions:
\textbf{Which intermediate representations are required to capture the properties of a physical system, yet allow for automatic transformations, optimisations, and scheduling to target specific hardware constraints?}
and
\textbf{How can noise models, error mitigation, and correction strategies be formally integrated into quantum software toolchains, such that their influence on performance can be systematically measured?}
Formal abstractions, intermediate representations for digital \emph{and} analogue operations, and tools for automated mapping and optimisation across hardware platforms are required.
Furthermore, quantum simulation spans a spectrum from fully digital (gate-based) to fully analogue approaches.
No systematic framework to evaluate or compare different simulation strategies along this continuum is available.
This lack of software support limits the ability to make informed choices about how to implement a given simulation task.
We therefore ask:
\textbf{How can software architecture and patterns help select among digital, analogue, and hybrid simulation strategies based on the characteristics of the model and available hardware?}
and
\textbf{What benchmarking tools and performance models are needed to evaluate quantum simulation implementations--including trade-offs between quantum and classical backends, and across the digital-analogue spectrum?}
Addressing these challenges requires not only new benchmarking methodologies but also software that can model, simulate, and optimise across the full
design space.

\begin{acks}
This work is supported by the German Federal Ministry of Research, Technology and Space (BMFTR), funding program \emph{quantum technologies---from basic research to market}, grant 13N17387, by the High-Tech Agenda of the Free State of Bavaria, and by the German Research Foundation, grants MA9739/1-1 and SCHA1635/20-1.
\end{acks}

\renewbibmacro{in:}{}
\renewbibmacro{doi:}{}
\newbibmacro{string+doi}[1]{%
  \iffieldundef{doi}{\iffieldundef{url}{#1}{\href{\thefield{url}}}{\textcolor{teal}{#1}}}%
  {\href{https://dx.doi.org/\thefield{doi}}{\textcolor{teal}{#1}}}}
\DeclareFieldFormat*{title}{\usebibmacro{string+doi}{\mkbibemph{#1}}}

\printbibliography

\end{document}